\documentclass[letterpaper,10pt]{article}
\usepackage{epsfig}
\usepackage{graphicx}
\usepackage{longtable}
\usepackage{amssymb,amsmath}
\usepackage{natbib}
\usepackage{txfonts}
\usepackage{sidecap}
\usepackage{multicol}
\usepackage[pdftex]{hyperref}   % doit toujours tre chargŽ en dernier (sauf geometry): redŽfinit certaines commandes
\hypersetup{bookmarks=true,colorlinks=true,linkcolor=black,citecolor=blue,urlcolor=blue}

\newcommand{\aap}{A\&A}

\newcommand{\apjl}{ApJL}

\begin{document}

\title{ {\large Answering the {\it Spitzer} white paper call to open new proposal opportunities\thanks{\href{http://ssc.spitzer.caltech.edu/warmmission/sus/mlist/archive/2013/msg005.txt}{http://ssc.spitzer.caltech.edu/warmmission/sus/mlist/archive/2013/msg005.txt}
}} \\ A search for rocky planets transiting brown dwarfs}

%\author{  Amaury H.M.J. Triaud$^{1,15}$, Micha\"el Gillon$^2$, Franck Selsis$^3$, Joshua N. Winn$^1$, Gregory P. Laughlin$^4$\and
%\'Etienne Artigau$^5$, Thierry Forveille$^6$,  Michel Mayor\and
%\'Emeline Bolmont, Janis Hagelberg, J\'er\'emy Leconte, Monika Lendl, Stuart Littlefair, Sean Raymond, Johannes Sahlmann  ...
%}

\date{}
\maketitle

\vspace{-0.6in}
{ \centering Amaury H.M.J. Triaud$^{1,12}$, Micha\"el Gillon$^2$, Franck Selsis$^3$, Joshua N. Winn$^1$, Brice-Olivier Demory$^4$, 
\'Etienne Artigau$^5$,  Gregory P. Laughlin$^6$, Sara Seager$^4$, Christiane Helling$^7$, Michel Mayor$^8$,
Lo\"ic Albert$^5$, 
Richard I. Anderson$^8$,  \'Emeline Bolmont$^3$, Ren\'e Doyon$^5$, 
Thierry Forveille$^9$, Janis Hagelberg$^8$, J\'er\'emy Leconte$^{10}$, Monika Lendl$^8$, Stuart Littlefair$^{11}$, Sean Raymond$^3$, Johannes Sahlmann$^8$
}\\

%\vspace{0.2in}
\footnotesize{  
\indent 1. Kavli Institute for Astrophysics \& Space Research, Massachusetts Institute of Technology, Cambridge, MA 02139, USA\\
\indent 2. Institut d'Astrophysique et G\'eophysique, Universit\'e de Li\`ege, all\'ee du 6 Ao\^ut 17, B-4000 Li\`ege, Belgium\\
\indent 3. Universit\'e de Bordeaux, Observatoire Aquitain des Sciences de l'Univers, BP 89, 33271 Floirac Cedex, France\\
\indent 4. Department of Earth, Atmospheric and Planetary Sciences, Massachusetts Institute of Technology, Cambridge, MA 02139, USA\\
\indent 5. D\'epartement de Physique and Observatoire du Mont-M\'egantic, Universit\'e de Montr\'eal, C.P. 6128, Succ. Centre-Ville, Montr\'eal, QC H3C 3J7, Canada\\
\indent 6. UCO/Lick Observatory, Department of Astronomy \& Astrophysics, University of California at Santa Cruz, Santa Cruz, CA 95064, USA\\
\indent 7. School of Physics \& Astronomy, University of St Andrews, North Haugh, St Andrews, KY16 9SS, Scotland, UK\\
\indent 8. Observatoire Astronomique de l'Universit\'e de Gen\`eve, Chemin des Maillettes, 51, CH-1290 Sauverny, Switzerland\\
\indent 9. UJF-Grenoble 1/CNRS-INSU, Institut de Plan\'etologie et d'Astrophysique de Grenoble (IPAG) UMR 5274, 38041, Grenoble, France\\
\indent 10. Laboratoire de M\'et\'eorologie Dynamique, Institut Pierre Simon Laplace, CNRS, 4 place Jussieu, BP99, 75252, Paris, France\\
\indent 11. Department of Physics and Astronomy, University of Sheffield, Sheffield S3 7RH, United Kingdom\\
\indent 12. Fellow of the Swiss National Science Foundation\\
}
\vspace{0.1in}

%In the past two decades, humankind went on from discovering the first planets around other stars to studying and comparing the detailed properties of hundreds of extrasolar planets. 
%There are 
%now close to 800 planets listed on the Exoplanet Encyclopaedia (Schneider et al. 2011) in addition to  2\,300+ candidates announced by NASA's \textit{Kepler} mission 
%(Borucki et al. 2011, Baltalha et al. 2012). Those discoveries have brought 
 %the plurality of worlds to the fore-front of scientific discussion, for what has been uncovered is the vast variety that planetary systems  present: 
%a diversity in mass, radius, density, orbital architecture, eccentricity, atmospheric species, temperature profile... The exploration of that diversity, and the 
%study of planets and systems unlike anything that exists in our planetary system is answering  questions about the formation of our own solar system, and our position within 
%all the possible outcomes that Nature can produce. Those discoveries are changing humanity's picture of the universe and are touching at the self-conception of our societies. 
\small

Exoplanetary science has reached a
historic moment. The {\it James Webb Space Telescope} will be capable of probing the atmospheres of rocky planets, and perhaps even search for biologically produced gases.
However this is contingent on identifying suitable targets before the end of the mission. A race therefore, is on, to find transiting planets with the most favorable properties, in time for the launch. 
%through ground-based surveys of late M dwarfs (such as the MEarth survey), and space-based surveys around a broader range of stars (such as the recently-approved NASA {\it TESS} mission).

Here, we describe a realistic opportunity to discover extremely favorable targets -- rocky planets transiting nearby brown dwarfs -- using the {\it Spitzer Space Telescope} as a survey instrument. 
Harnessing the continuous time coverage and the exquisite precision of \textit{Spitzer} in a 5,400 hour campaign monitoring nearby brown dwarfs, 
we will detect a handful of planetary systems with planets as small as Mars.
% brown dwarfs and catch the transits of planets as small as Mars. 
%In August 2012 we submitted a proposal with such an aim and asked for 5,400 hours of observing time to be spread on several cycles.
The survey we envision is a logical extension of the immense progress that has been realized in 
the field of exoplanets and a natural outcome of the exploration of the solar neighborhood to map where the nearest habitable rocky planets are located (as advocated by the 2010 Decadal Survey).~Our program represents an essential step towards the atmospheric characterization of terrestrial planets and carries the compelling promise of studying the concept of {\it habitability}  
beyond Earth-like conditions. In addition, our photometric monitoring will provide invaluable observations of a large sample of nearby brown dwarfs situated close to the  M/L transition.
This is why, we also advocate an immediate public release of the survey data, to guarantee rapid progress on the planet search and provide a treasure trove of data for brown dwarf science.

%:  planets transiting brown dwarfs are the most favorable 
%for which it will be possible to reliably detect atmospheric features using the upcoming James Webb Space Telescope. 
%Future observations of our discoveries  
%smay provide humankind with a first attempt at the detection of another genesis.
%The same data will be used to study weather patterns on brown dwarfs and, by stacking the images, we expect to directly image gas giant companions. Our survey is of utmost interest 
%for several research fields and intimately links the understanding of brown dwarfs with planet formation and stellar formation and evolution.%, in one 

	\paragraph{a characterizable rocky exoplanet to turn JWST to\\}

	The study of exo-atmospheres is a fascinating and fast growing field, so far mostly restricted to close-in gas giants \citep{Seager:2010kx}. We propose to extend this highly important field
	to terrestrial planet atmospheres. %and enter an age where the detection of extraterrestrial life would be possible. 
%	From observations during a planetary occultation, we can reconstruct the planet's emission spectrum using spectro-photometry (eg. \citet{Anderson:2013fk}). 
	It is now widely recognized that M dwarf stars are attractive targets because of their small sizes, and consequently reduced size contrast between planet and star. 
	Even so, observing the atmospheres of any Earth-sized transiting planets around M dwarfs may be difficult with {\it JWST}.
	\citet{Belu:2011qy} calculated the expected signal to noise and the necessary photometric precision associated with the detection of atmospheric features
	%\footnote{defined as a brightness temperature variation of 30~K between two adjacent wavelength bins.} 
	via transmission and reflection spectroscopy using MIRI (on board {\it JWST}). Reaching a $5\sigma$ detection of spectral features
	in the atmosphere of a rocky planet in the habitable zone of an M dwarf will require co-adding together almost every single occultation and transit occurring during {\it JWST}'s entire lifetime. 
%	In addition to those scheduling concerns, there are currently no surveys able to deliver an Earth-sized planet whose atmosphere would be 
%	detectable using {\it JWST}.
	
\begin{figure*}[t]
	\centering
	\includegraphics[width=12.cm, angle=0]{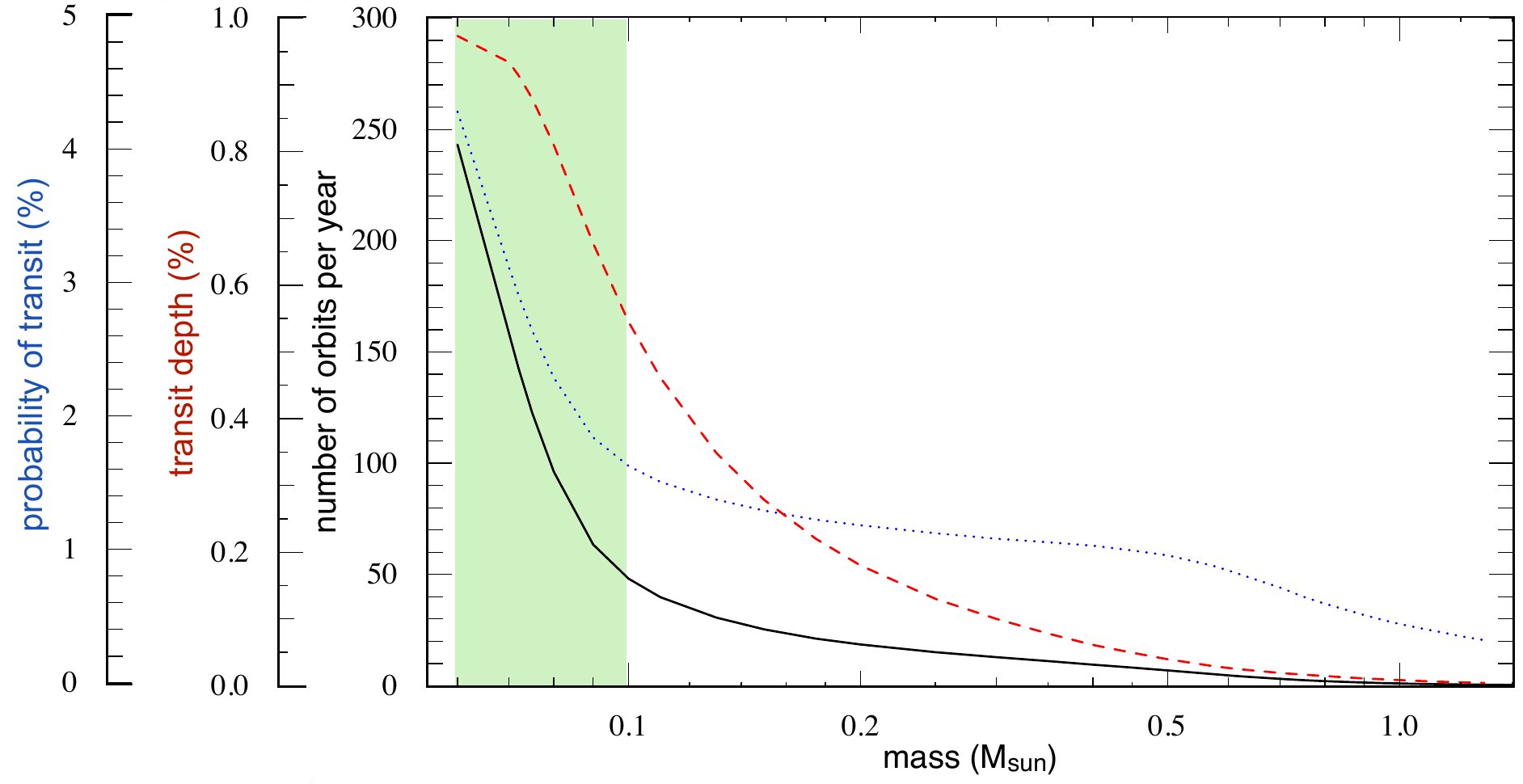}
	\caption{ \small{If you like M dwarfs, you'll love brown dwarfs.
	For a given equilibrium temperature (here 255K, like Earth), the number of orbits (i.e. transits or occultations) per year (black), 
	the transit depth (dashed red), and the probability of transit (dotted blue) 
	as a function of the primary's mass. The green area is where we plan our survey.
	Stellar parameters were obtained from a 1~Gyr isochrone \citep{Baraffe:1998ly}. The natural evolution of stars and brown dwarfs will alter those curves with time; for brown dwarfs, 
	the rising slopes get steeper.}
	}\label{BDs}
	\end{figure*}

\begin{figure*}[t]
	\centering
	\includegraphics[width=7.4cm, angle=0]{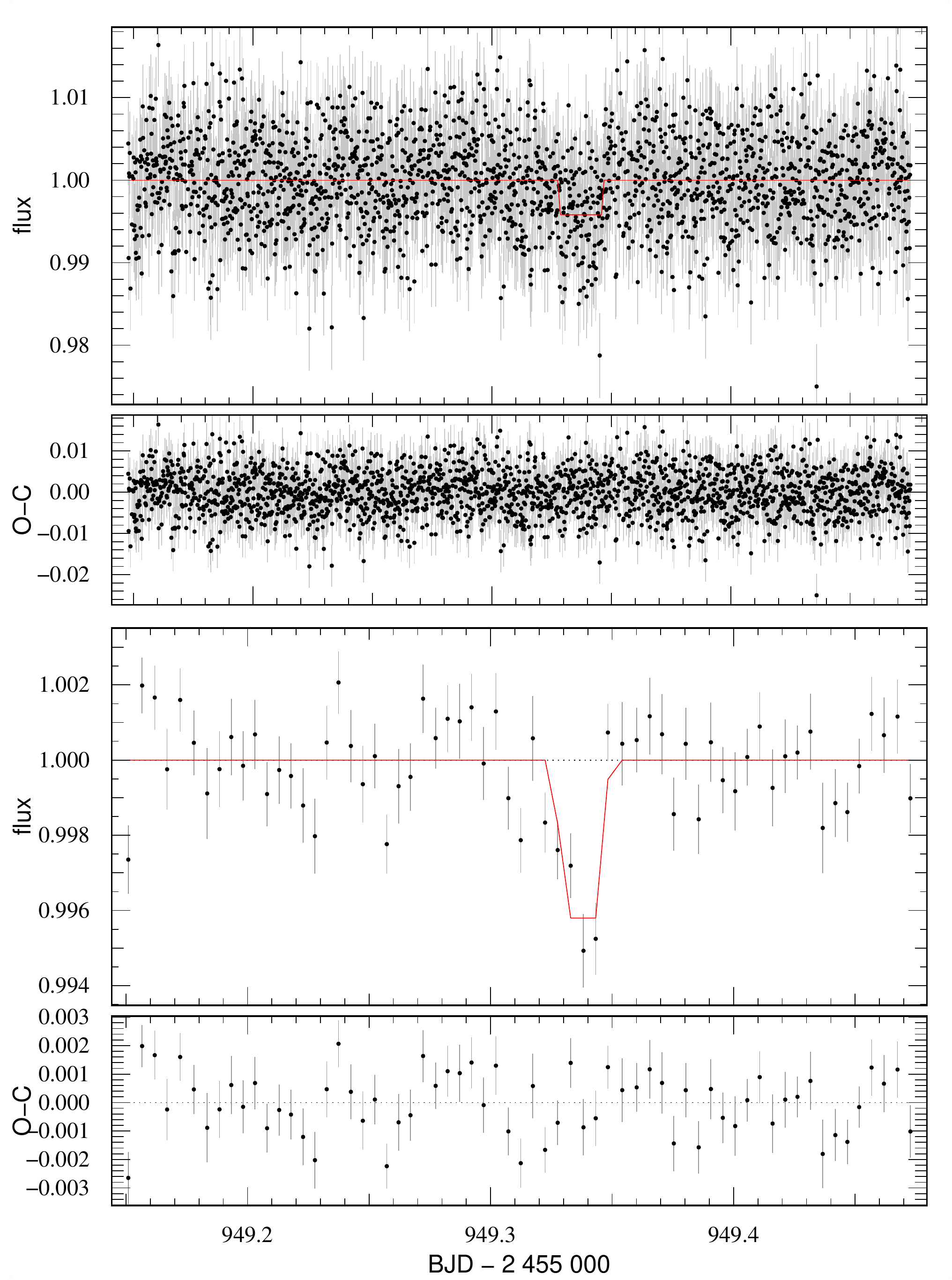}
	\includegraphics[width=7.4cm, angle=0]{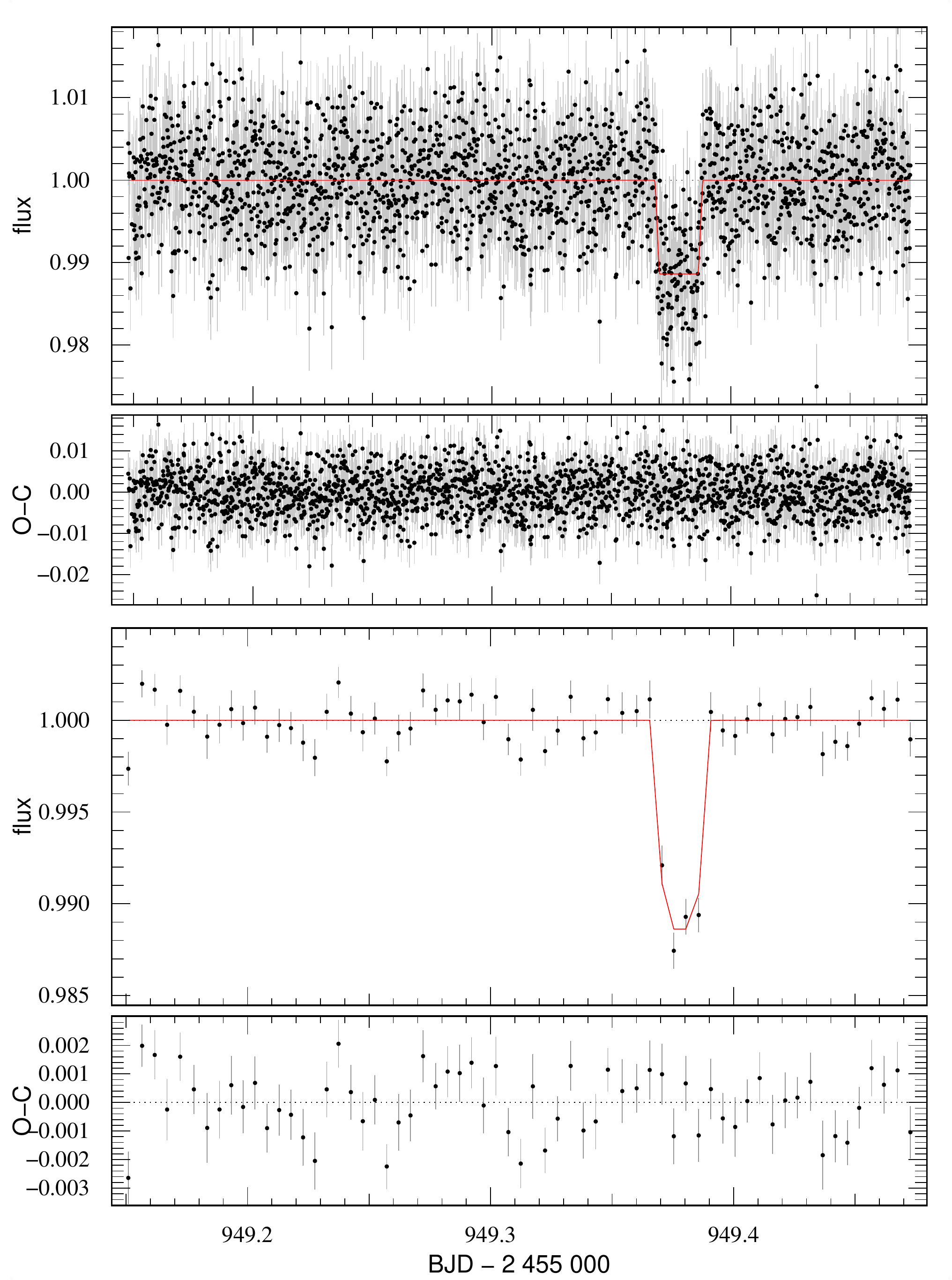}
	\caption{\small{\textbf{top} archival \textit{Warm Spitzer} [4.5 $\mu$m] data superposed with a simulated transit. We fitted a transit model and included corrections for systematics 
	(notably intra-pixel variations). \textbf{bottom} data and model 
	have been binned for visual clarity. \textbf{left} a 1~R$_{\rm mars}$ planet transiting a 0.9~R$_{\rm Jup}$ object \citep{Triaud:2013lr}. \textbf{right}, a 1~R$_{\rm earth}$. The transits are 
	located at different times. The residuals (O-C) are similar, illustrating our ability to correct known systematics, distinguish the signal and not over-fit.}}\label{transitsimul}
	\end{figure*}

	We extended this work in the context of brown dwarf primaries who present more favorable characteristics for the detection but also for the future characterization of rocky exoplanets
	(see Fig. \ref{BDs}).
	When wishing to detect photons emitted by an exo-atmosphere (at occultation), the distance of that exoplanet relative to the Solar System is 
	what primarily matters. In addition, for a given planetary equilibrium temperature, the contrast between the central object 
	and the planet is highest and most favorable, the fainter the primary. This makes brown dwarfs natural targets to consider; they possess other advantages: 
	\begin{itemize} \itemsep0em
	\item{for a given planetary equilibrium temperature, the orbit gets shorter with decreasing primary mass, 
	increasing the probability of transit and providing 50+ occultations per year (and 50+ transits) (Fig. \ref{BDs}); } 
	\item{the planet to brown dwarf size ratio means transiting rocky 
	planets produce deep transits and permit the detection of planets down to Mars' size in a single transit event when using {\it Spitzer} (Fig. \ref{transitsimul});}
	\item{the reliability of the detection is helped by the absence of known false astrophysical positives: brown dwarfs have very peculiar colors, small sizes, and being nearby, have 
	a high proper motion allowing to check what is within their glare.}
	\end{itemize}
	 Brown dwarfs older than $\sim$ 0.5 Gyr have a near constant radius over their mass range: a fairly accurate estimation of the size 
	of the planet can be obtained without requiring a complete characterization of the host as when we search for transits on solar-like stars. 
	
	Observing in broad band would reveal the presence of an atmosphere through the phase curve 
	\citep{Maurin:2012yq}. Using narrow bands can provide spectral signatures \citep{Selsis:2011lr}. The amplitude of the phase curve of a planet in the habitable zone, using 
	ten bands with MIRI can be detected at $10 \sigma$, by observing continuously during two orbital periods.  A $5\sigma$ detection of spectral features in emission
	will be reached in just 50 occultations (within a year)\footnote{simulation only realized for MIRI, but NIRISS, NIRcam and NIRSpec, also onboard {\it JWST}
	can study the atmospheres of the planets we will find.
	%the slitless mode of NIRISS (also on {\it JWST}) is also considered for such observations \citep{Doyon:2012lr}.
	}.
	This jump in sensitivity is primarily caused by the fast occurrence of occultations,
	thanks to short habitable orbital periods (see also \citet{Belu:2013vn}).
	
	We will not enter here into considerations about habitability nor about the chemical and photo-chemical processes leading to the eventual emergence of observables
	 biomarkers in a planet's spectrum. 
	We simply note that 
	any discovered planet will become a prime laboratory to study exo-atmospheres and will form a benchmark against which any future mission will refer itself too. 
	\textit{Planets transiting brown
	dwarfs  offer the fastest and most convenient route to the detection and to the study of the atmospheres of terrestrial extrasolar planets.} 

	\setlength{\textfloatsep}{5pt}

	\paragraph{the existence of planets orbiting brown dwarfs\\}
	
	Searching for planets around brown dwarfs is close to virgin territory. We can nevertheless build on a few discoveries to extrapolate to this new region of parameter space.
	The MEarth project found a 6.6~M$_{\rm earth}$ planet transiting a 0.16~M$_\odot$ star \citep{Charbonneau:2009fj}. 
%	Sadly, MEarth will not be able to detect planets smaller than 1.5~R$_{\rm earth}$ (Berta et al. submitted). 
	There are two microlensing events on low mass stars reported in the literature  including one 
	caused by a 3.2~M$_{\rm earth}$ orbiting a primary at the limit between stars and brown dwarfs with a mass of only 0.084~M$_\odot$ \citep{Kubas:2012fk}.
	Accounting for their low probabilities, such detections indicate the presence 
	of a large, mostly untapped, population of low mass planets around very low mass stars (see also \citet{Dressing:2013fj}).
	Arguably the most compelling discovery is that of the \textit{Kepler} Object of Interest 961, a 0.13~M$_\odot$ star, orbited by a 0.7, a 0.8 and a 0.6~R$_{\rm earth}$ on periods shorter 
	than two days \citep{Muirhead:2012fk}. The KOI-961 system, remarkably, appears like a scaled-up version of the Jovian satellite system. This is precisely what we are looking for. 

	Disks around brown dwarfs can be relatively large ($>$ 10 AU), massive and long-lived, providing in principle sufficient mass for the formation of gas giants 
	(eg. \citet{Scholz:2008kx}). While models show gas giants are not expected, smaller mass planets are \citep{Payne:2007lr}.
	Planet formation models also indicate the existence of "bottlenecks" to growth due to dynamical and hydrodynamical processes.  
	This leads to a pile-up of  "planetary embryos" with typical masses of order 1~M$_{\rm mars}$ (0.1 M$_{\rm earth}$) 
	\citep{Kokubo:2002uq}. Those embryos form the building blocks of larger rocky planets such as Earth. Our survey is meant to be able to detect such small planets.
	While being able to detect Mars-sized objects may increase our chance of detecting any system around brown dwarfs, their presence, or their absence, will have a direct feedback 
	on the planet formation models themselves. 
	
	\setlength{\textfloatsep}{10pt}
	
	\paragraph{detecting rocky planets transiting brown dwarfs\\}
	Our aim is to be able to detect a single transit event.
	Observations carried out on \textit{Spitzer} during Cycle 8 (e.g. \citet{Heinze:2013uq}) have shown that whereas photon noise is slightly better in 
	channel 1 at [3.6\,$\mu$m] than at [4.5 \,$\mu$m], the amplitude of the intra-pixel variability and the number of systematics are smaller in the latter. 
	We simulated a Mars-sized transit event by inserting the signal we seek on real archival \textit{Spitzer} data of a typical brown dwarf within our sample (a J=13.7, L3 dwarf). 
	The study of those simulations indicates that variability in channel 1  can sometimes hide a transit signal, or produce features that can be confused for transits.
	A single transit detection of a Mars-sized planet cannot be secured when observing at [3.6\,$\mu$m]. This motivates our choice in favor of  channel 2 
	where similar simulations showed we can distinguish a single planet's signal unambiguously (see Fig. \ref{transitsimul}).

	We used the number of dwarf stars as red, or redder than KOI-961 observed by {\it Kepler} to infer that $\sim$ 40\% of objects 
	have a planetary system with short orbits\footnote{A {\it Kepler} GO proposal focusing on 
	1200 late M dwarfs (PI Demory) will soon confirm this occurrence
	of planetary systems close to the brown dwarf range}. 
	Placing a planet on a 30 hour orbit (KOI-961b), it has a 4\% probability to transit. Those simple numbers indicate that to 
	obtain two detections, it is required to observe 120 targets for 30 hours continuously, a program close to 4,000 hours of 
	effective observations and overheads. More detailed simulations revealed that   a minimum of 5,400 hours of {\it Spitzer} time is necessary to yield
	a 90\% probability of finding one or more transiting systems. Our most pessimistic case gives a 60\% chance of success, while our most optimistic predicts a 
	10\% probability to discover 10 or more systems (with the chance of each containing several planets). 
	
	As described, our survey is limited only by {\it Spitzer}'s observing time and 
	not by the availability of bright-enough brown dwarfs. It can be extended beyond 5,400 hours in the future, to increase the probability of a detection.

%	This is also favorable for the direct imaging side of our time request since this band is the closest to the peak 
%	of the black-body radiation emitted by Jovian planets. By offering a smaller contrast this gives us \textit{the opportunity to search for lower mass, colder, companions} 
%	which have remained hidden in the existing surveys.

	\paragraph{summary\\}
	
	The recent selection of {\it TESS} by NASA testifies to the intense interest in the discovery of nearby transiting planets whose atmospheres can be more readily studied.	
%	The search for transiting rocky planets whose atmospheres will be characterizable using the JWST is one of the most exciting science goals and one that has been
 %	recently supported by the selection of TESS.
	By surveying brown dwarfs, {\it Spitzer} has a chance of finding even more favorable objects for atmospheric spectroscopy.
	Certainly, there are uncertainties regarding planets orbiting brown dwarfs and their possible habitability, that do not exist to the same degree for normal stars. Yet, past exoplanet discoveries 
	have taught us that we should observe without too much theoretical prejudice. Our program will also open the door to the study of planet formation processes at the 
	very bottom of the main sequence.
	
	{\it Spitzer} is the only facility that can 
	survey a sufficient number of brown dwarfs, long enough, 
	with the precision and the stability required to credibly be able to detect rocky planets down to the size of Mars, in time for {\it JWST}. We 
%	We will then turn to large ground-based facilities to confirm the transit, find the period (if only one event has been captured) and check for the presence of additional companions. 
%	Carrying a similar survey on those same telescopes
%	would require about five years of near continuous use and locations in both hemispheres, while having to deal with the known contingencies about observing in 
%	the near IR through the atmosphere and having the day/night cycle affecting the window function. Ground-based observations remain limited to a small number 
%	of  close and bright objects such as Luhman-16 \citep{Luhman:2013lr,Gillon:2013qy}.
	estimate that about 8 months of observations would be needed to complete the survey\footnote{Scheduling constraints would be minimal since any system can be targeted at any time; 
	the observing load could be spread over several cycles. Thanks to the intrinsic faintness of our targets, it is also a program requiring low volumes of downlink.}.
	Once candidates are detected, large ground-based facilities will confirm the transits, find the period (if only one event was captured by {\it Spitzer}) 
	and check for the presence of additional companions.
%	ensures the right precision and thus, the reliability of discovering a planet transiting a nearby brown dwarf. 
%	The only requirement is that the series be taken as continuously as permitted currently by the spacecraft. 
	This program will rapidly advance the search for potentially habitable planets in the solar neighborhood and transmit to {\it JWST} a handful of characterizable rocky planet atmospheres.	
	
	Because of the wide interest in the discovery of a nearby transiting Earth-like planet, and the time pressure to identify such targets in time for {\it JWST} observations, we 
	advocate for publicizing the survey data immediately and thereby allowing anyone to become involved. 
	 We also anticipate that the variability data will be of great interest to the brown dwarf community.
		
%	We are aware that this venture is somewhat risky, but the possible gains are also very significant. 
%	We would like to take advantage that it is nearing the end of its life and 
%	is getting further from Earth to obtain those observations and transmit to JWST the strong legacy of a handful of characterizable rocky planet atmospheres.% with its enormous scientific and philosophic impact.% and we propose to spread the observations on several cycles to prevent monopolizing the telescope.
%	
%	\cite{1995Natur.378..355M}

%	We welcome the exoplanet and the brown dwarf communities to join our efforts to carry out this search.
	
%	Finding such planets from the ground is possible, but only using medium to large size telescopes, situated in both hemispheres and surveying the sky for about five years.
%\bibliography{bibtex}

%\begin{thebibliography}

%\bibitem{1995Natur.378..355M} Mayor, M., \& Queloz, D.\ 1995, Nature, 378, 355 

%\end{thebibliography}
\vspace{-0.1in}
{\footnotesize
\bibliographystyle{aa}
\bibliography{../1Mybib}
}

\end{document}